\documentclass[reprint, amsmath,amssymb,aps,twocolumn, prb,superscriptaddress]{revtex4-1}
\usepackage{amsfonts,amssymb}
\usepackage[subnum]{cases}
\usepackage{mathrsfs}
\usepackage{amsmath}
\usepackage[none]{hyphenat}
\usepackage{graphicx}
\usepackage{hyperref}    
\usepackage{bm}          
\usepackage{natbib}
\usepackage{soul} 
\usepackage{color}
\usepackage{longtable}
\usepackage{ textcomp }
\usepackage{verbatim}

\begin{document}

\title{Surface roughness scattering in multisubband accumulation layers}

\author{Han Fu}
\email{fuxxx254@umn.edu}
\affiliation{Fine Theoretical Physics Institute, University of Minnesota, Minneapolis, MN 55455, USA}
\author{K. V. Reich}
\affiliation{Fine Theoretical Physics Institute, University of Minnesota, Minneapolis, MN 55455, USA}
\affiliation{Ioffe Institute, St Petersburg, 194021, Russia}
\author{B. I. Shklovskii}
\affiliation{Fine Theoretical Physics Institute, University of Minnesota, Minneapolis, MN 55455, USA}
\date{\today}

\begin{abstract}
Accumulation layers with very large concentrations of electrons where many subbands are filled became recently available due to ionic liquid and other new methods of gating. The low temperature mobility in such layers is limited by the surface roughness scattering. However theories of roughness scattering so far dealt only with the small-density single subband two-dimensional electron gas (2DEG). Here we develop a theory of roughness-scattering limited mobility for the multisubband large concentration case. We show that with growing 2D electron concentration $n$ the surface dimensionless conductivity $\sigma/(2e^2/h)$ first decreases as $\propto n^{-6/5}$ and then saturates as $\sim(da_B/\Delta^2)\gg 1$, where $d$ and $\Delta$ are the characteristic length and height of the surface roughness, $a_B$ is the effective Bohr radius. This means that in spite of the shrinkage of the 2DEG width and the related increase of the scattering rate, the 2DEG remains a good metal. Thus, there is no re-entrant metal-insulator transition at high concentrations conjectured by Das Sarma and Hwang [PRB 89, 121413 (2014)].
\end{abstract}

\maketitle

\section{Introduction}
The electron mobility is a very important parameter of electronic devices. In heavily doped bulk semiconductors, the low temperature mobility is determined by the electron scattering on ionized donors and is relatively small. Larger low temperature mobilities can be achieved near the surface of a lightly doped say $n$-type semiconductor, where the electron accumulation layer is induced by the applied surface electric field. In such devices the  mobility becomes sensitive to the semiconductor surface roughness, which can be imagined as a collection of atomic-size steps of total height $\Delta$ and characteristic size $d \gg \Delta$ along the surface. The roughness scattering dominates at high electric fields $E$ when electrons are squeezed closer to the surface. In this case~\cite{Prange,Chaplik,*Entin,Ando_1977,Ando_1979,Ando_1982,Suris} the mobility $\mu$ limited by the surface roughness scattering behaves as $\mu \propto 1/E^{2}$. For a large enough field $E$ the two-dimensional (2D) concentration of electrons $n \propto E$ so that $\mu \propto 1/n^{2}$  and the surface conductivity $\sigma = ne\mu \propto 1/n $. This result holds for an inversion layer in a lightly doped $p$-type semiconductor when the electric field $E$ is larger than the electric field of the depletion layer. The low temperature mobility was also extensively studied in quantum wells, where it is limited by the surface roughness of both interfaces. This mobility strongly depends on the width of the quantum well~\cite{Sakaki, Gold, Penner}.

Because of the interest in higher mobilities, the surface roughness scattering was studied theoretically only for relatively small concentrations $n$, when only the first energy quantization subband is filled at low temperatures~\cite{Chaplik,*Entin,Ando_1977,Ando_1979,Ando_1982,Suris,Sakaki, Gold, Penner}. Also, it was difficult to induce large electron concentrations $n$ (higher than $10^{13} ~\mathrm{cm}^{-2}$ in Si). So the the roughness scattering in the case of large concentrations $n$ when many subbands are filled at low temperatures has remained unexplored.

The last decade, however, witnessed growing interest in accumulation layers with large $n$ which allow to achieve qualitatively new properties of the electron gas, such as superconductivity or magnetism. New methods to create large electron concentrations were developed. One of them is based on ion gating with help of an electrolyte or a room temperature ionic liquid, which does not need an insulator layer and, therefore, makes a double layer with a very large capacitance. In Si concentrations $n \sim 5 \cdot 10^{13}~\mathrm{cm}^{-2}$ were achieved using gating by an electrolyte~\cite{Tardella} and by an ionic liquid~\cite{JJ}. Even larger concentrations $\sim 10^{14}~\mathrm{cm}^{-2}$ were induced in ZnO~\cite{Iwasa_2009}, $\mathrm{MoS_2}$~\cite{Iwasa_2012} and $\mathrm{SrTiO_3}$ \cite{Iwasa_2008,Gallagher_2014} with this method.

Another important method is based on heterojunctions of polar and nonpolar perovskites such as $\text{GdTiO}_3$ and $\text{SrTiO}_3$, which accumulate $3 \cdot 10^{14}~\mathrm{cm}^{-2}$ electrons \cite{Stemmer}. Concentrations $n$ up to $10^{15}~\mathrm{cm}^{-2}$ were obtained combining this effect with the electron spill-out for a special band alignment  \cite{Bharat}. Similar physics takes place in GaN-based heterojunctions where concentrations up to $4.4\times10^{13}$ cm$^{-2}$ were achieved \cite{Shur, Speck_2009, Shur_2009}.

At such large concentrations $n$ the dimensionless parameter $na_{B}^{2} > 1$  and many electron subbands are filled. Here  $a_B=\kappa\hbar^2 /m^*e^2$ is the effective Bohr radius, $\kappa$ is the dielectric constant, $m^*$ is the effective electron mass. In the cases of ZnO and MoS$_2$ mentioned above $na_{B}^{2}$ reaches 5. In semiconductors with relatively large $a_{B}$ such as GaAs, InAs, InSb, and PbTe, it should be easy to reach  $na_{B}^{2} \gg 1$.

\begin{figure}[t]
 \includegraphics[width=.45\textwidth]{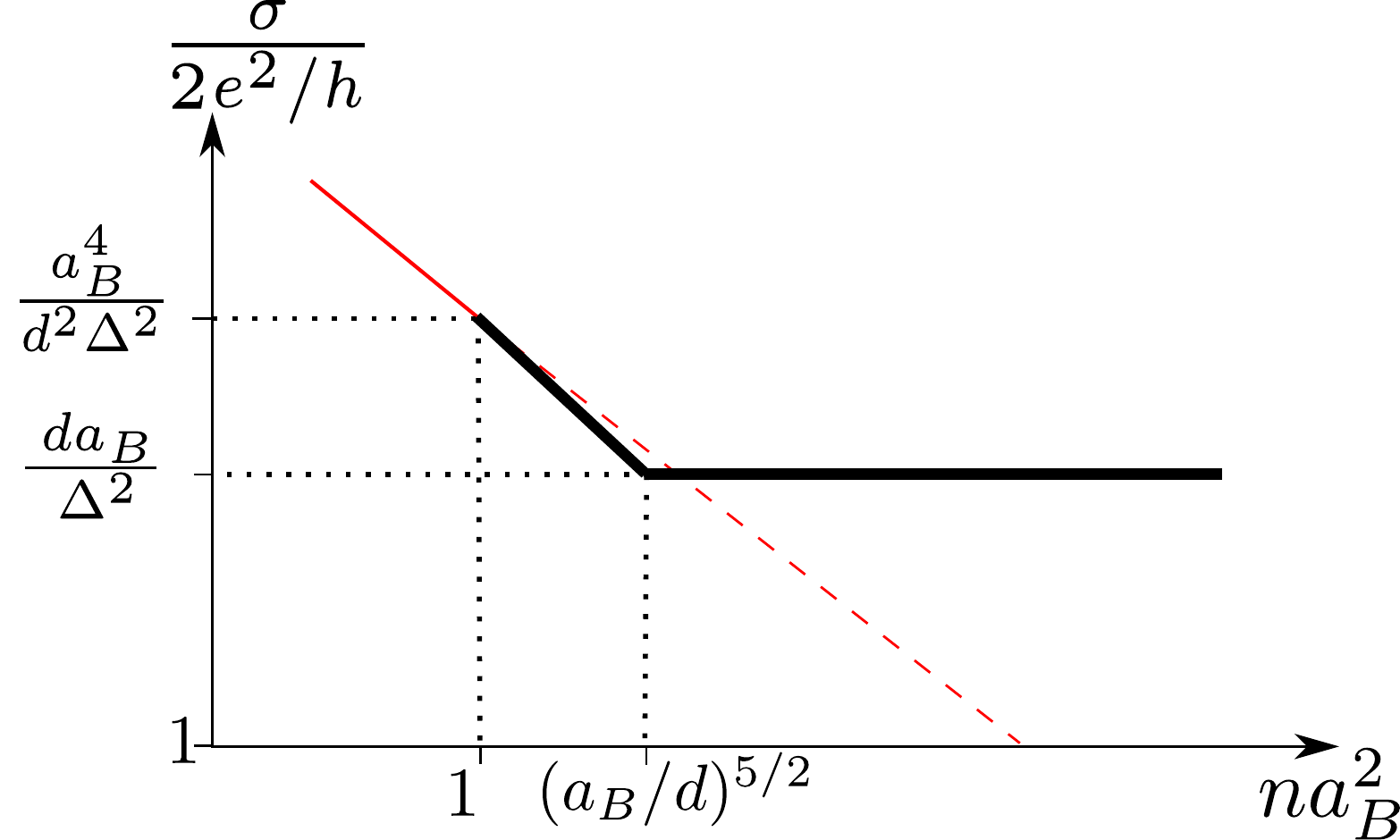}\\
 \caption{(Color online) The schematic log-log plot of the dimensionless conductivity of an accumulation layer $\sigma /(2e^2/h)$ limited by the surface roughness scattering as a function of the dimensionless 2D electron concentration $na_{B}^{2}$ at $d<a_B$.  The thick solid line (black) shows the conductivity for the multisubband accumulation layer. It first decreases as $(na_B^2)^{-6/5} (a_B^4/d^2\Delta^2)$ and then saturates at $na_B^2\sim (a_B/d)^{5/2}$, where the wavelength $k_F^{-1}\sim d$. The thin solid line (red) represents the $1/n$ dependence derived for a single subband \cite{Ando_1977}. Conjectured extrapolation \cite{DasSarma} of this dependence to larger  concentrations is shown by the thin dashed thin (red).}\label{fig:conductivity_the_result}
 \end{figure}

As we said above the roughness scattering limited mobility for the multisubband case has not been theoretically studied. In this paper, we fill this gap and study the low temperature mobility limited by surface roughness in an accumulation layer with large $n$. Our result for the dimensionless conductivity $\sigma(n) /(2e^2/h)$ at $d<a_B $ is shown on Fig. \ref{fig:conductivity_the_result} as a function of the dimensionless concentration  $na_{B}^{2}$ for the exponential model of roughness \cite{Goodnick} by the thick solid line (black). The conductivity first decreases with $n$ as $\propto n^{-6/5}$ and then saturates at the level
\begin{equation}\label{eq:conductivity}
\frac{\sigma}{2e^2/h}\simeq\frac{da_B}{\Delta^2},
\end{equation}
which is much larger than unity assuming that both $d\gg \Delta$ and $a_B\gg \Delta$. The thin solid line (red) schematically shows the $1/n$ dependence of the conductivity derived for a single subband by previous work~\cite{Ando_1977}.

We see that with growing $n$ our conductivity at first approximately continues the single subband dependence $1/n$, but then saturates. The saturation happens when the electron wavelength $k_F^{-1}$ is equal to the size $d$ of the roughness. Before this point, the roughness felt by electrons is averaged over all irregularities within the region of size $k_F^{-1}$. As the concentration increases, $k_F$ increases and fewer irregularities are averaged over making the surface ``rougher" for electrons. When $k_F^{-1}$ gets below $d$, the electron ``hits" only a single irregularity and the level of ``roughness" is fixed leading to the saturation of the conductivity.

Our results contradict to the conjecture~\cite{DasSarma} that the single subband result~\cite{Ando_1977} can be extrapolated to the large-concentration multisubband case (see the thin dashed line (red) in Fig. \ref{fig:conductivity_the_result}). This conjecture led to the dramatic prediction~\cite{DasSarma} that the dimensionless conductivity could become smaller than unity implying the re-entrant metal-insulator transition with growing $n$. Our results show that at large $n$ the accumulation layer remains metallic. This agrees with decent mobilities observed in Refs. \onlinecite{diamond,Iwasa_2009,Iwasa_2012,Tardella}.

The organization of our paper is as follows. In Sec. \ref{sec:accu}, we explain the structure of the accumulation layer where electrons occupy many energy quantization subbands. In Sec. \ref{sec:roughness}, we introduce two models of the surface roughness including the exponential one assumed in Fig. \ref{fig:conductivity_the_result} and the Gaussian widely used in earlier studies. In Sec. \ref{sec:class}, we present an intuitive quasi-classical interpretation of our mobility results for the exponential roughness. In Sec. \ref{sec:sing}, we introduce the more formal quantum-mechanical approach starting from the case of a single subband connecting to previous studies. In Sec. \ref{sec:inter}, we discuss the multisubband case, take into account the scattering of electrons between different subbands, and give the final scattering rate and mobility. We conclude in Sec. \ref{sec:con}. In the main text of the paper we use the scaling approach and drop numerical coefficients. In Appendix \ref{App:AppendixA} we estimate the coefficients of the conductivity for the exponential roughness.

\section{Electronic structure of an accumulation layer}\label{sec:accu}

\begin{figure}[h]
\includegraphics[width=.4\textwidth]{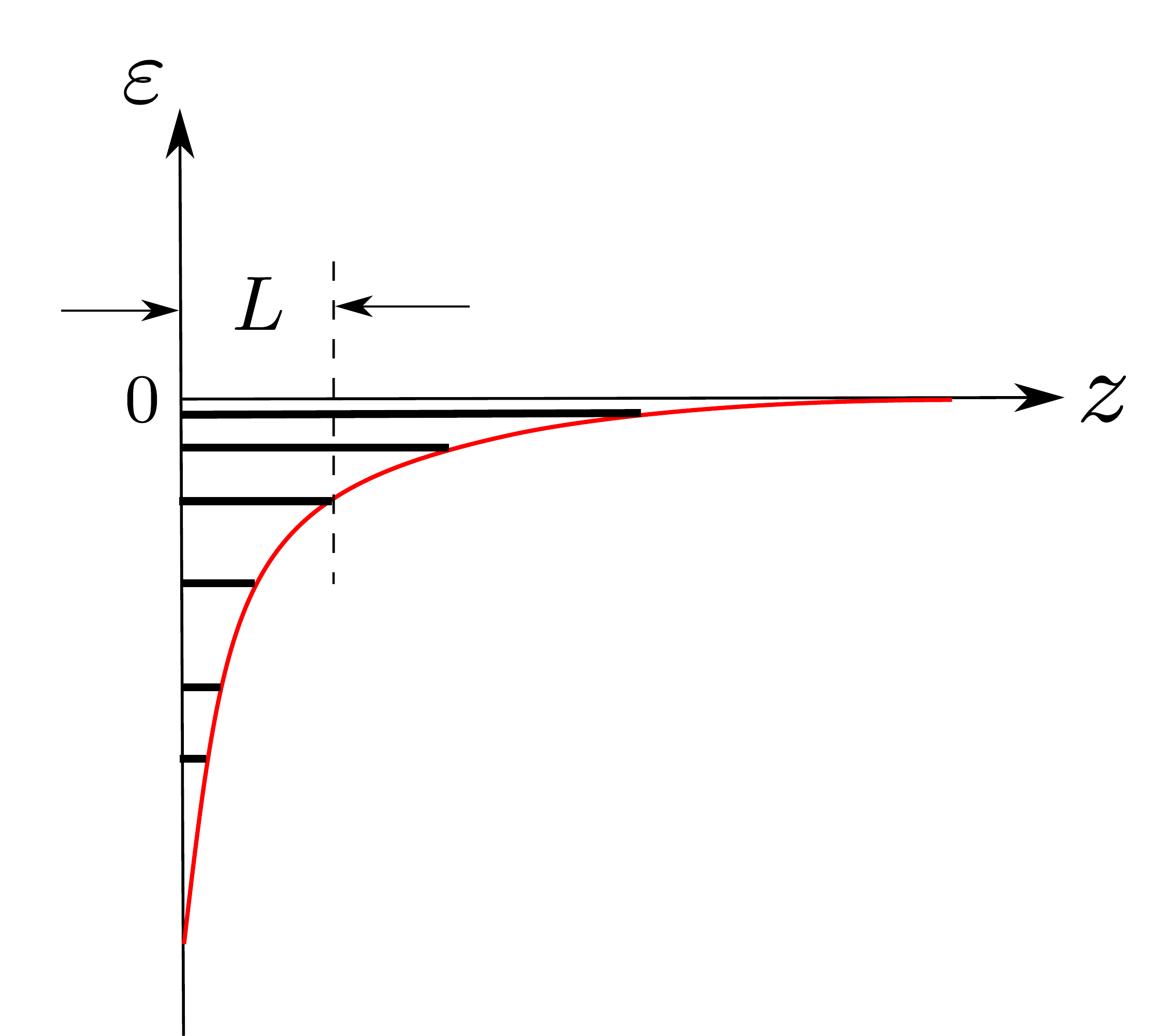}\\
\caption{(Color online) Schematic plot of the subbands electrons occupy at low temperatures in the accumulation layer. The subbands are represented by the black thick horizontal lines. The electron energy $\varepsilon$ consists of the kinetic energy $\hbar^2k^2/2m^*$ where $k$ is the momentum and the potential energy $-e\varphi(z)$. The surface potential well $-e\varphi(z)$ is shown by the grey thin line (red) where $\varphi(z)$ is given by Eq. \eqref{eq:potential_linear}. The reference energy level $\varepsilon=0$ is the conduction band bottom. In a lightly $n$-doped bulk semiconductor the electron Fermi level is close to zero. Most electrons are located within $z\lesssim L$ where $L$ is the decay length given by Eq. \eqref{eq:decay}.}\label{fig:subband}
\end{figure}

The accumulation layer is created near the surface of an $n$-type semiconductor when the orthogonal-to-surface electric field $E$ induces a large 2D concentration of excessive surface electrons $n=E/4\pi e$ in the layer of width $\sim L$. We assume that $n/L$ is much larger than the bulk concentration of electrons $N$ in the semiconductor. This means that the bulk Fermi level at low temperatures is either below the conduction band bottom or slightly above it. In Fig. \ref{fig:subband} illustrating the accumulation layer we actually assumed that the bulk Fermi level coincides with the conduction band bottom, which in this paper serves as the reference point of the electron energy. Our description of the electron accumulation layer is applicable also to an inversion layer in the very lightly doped $p$-type semiconductor, where the 2D concentration $n_{depl}$ of ionized acceptors forming the depletion layer is much smaller than the 2D concentration $n$ of the electrons so that we can still use $n=E/4\pi e$.

In this paper we are focused on large $E$ and $n$ cases when the one-dimensional potential well created by the electric field $E$ along the $z$-axis (which is defined as normal to the surface towards the accumulation layer) is so deep that it has several quantized levels (see Fig. \ref{fig:subband}). Each of such levels forms a subband with a two-dimensional (2D) Fermi gas moving freely parallel to the surface. At large $n$ all these 2D gases with the same Fermi level form a three-dimensional (3D) degenerate gas, which can be described in the Thomas-Fermi approximation neglecting the discreteness of subbands. The nonlinear screening of the electric field $E$ by such a gas was studied by solving the Poisson-Thomas-Fermi equation for the self-consistent potential $\varphi(z)$ and the 3D electron density $N(z)$ by Frenkel~\cite{Frenkel}. The result is~\cite{Frenkel,Kostya}
\begin{equation}
  \label{eq:potential_linear}
  \varphi(z)=  C_1 \frac{e}{\kappa a_B}  \left(\frac{a_B}{z+L} \right)^4,
\end{equation}
\begin{equation}
\label{eq:concentration_linear}
N(z)= C_2 \frac{1}{a_B^3} \left(\frac{a_B}{z+L}\right)^{6},
\end{equation}
where  $z$ is the distance from the interface, $a_B=\kappa\hbar^2/m^*e^2$ is the effective Bohr radius, $\kappa$ is the
dielectric constant, $m^*$ is the isotropic effective electron mass, $C_1=225\pi^2/8 \simeq 278$, $C_2=1125 \pi /8 \simeq 442 $. $L$ is the characteristic decay length of the electron density
\begin{equation}
\label{eq:decay}
L=C_3 \frac{a_B}{(na_B^2)^{1/5}},
\end{equation}
where $C_3=(225\pi/8)^{1/5} \simeq 2.4$, $n$ is the total 2D concentration of electrons inside the accumulation layer. The width of the electron gas is $\sim L$. This solution is valid for $na_B^2 \gg 1$, when for $N(0)\sim {n/L} \sim (na_B^2)^{6/5}{a_B}^{-3}$ and for $k_F \sim N(0)^{1/3} = (na_B^2)^{2/5}/a_B$ the following inequalities hold: $N(0)a_B^{3} \sim (na_B^2)^{6/5} \gg 1$ and $k_F L \sim (na_B^2)^{1/5} \gg 1$. These inequalities confirm that we deal with a 3D degenerate gas and that the Thomas-Fermi approximation is valid and many subbands are filled.
At smaller concentrations when $na_B^2 < 1$ only one subband is filled so that we go back to the 2D case studied for the inversion layer or narrow quantum wells.

\section{Models of surface roughness}\label{sec:roughness}
\begin{figure}[h]
$\begin{array}{c}
\includegraphics[width=0.47\textwidth]{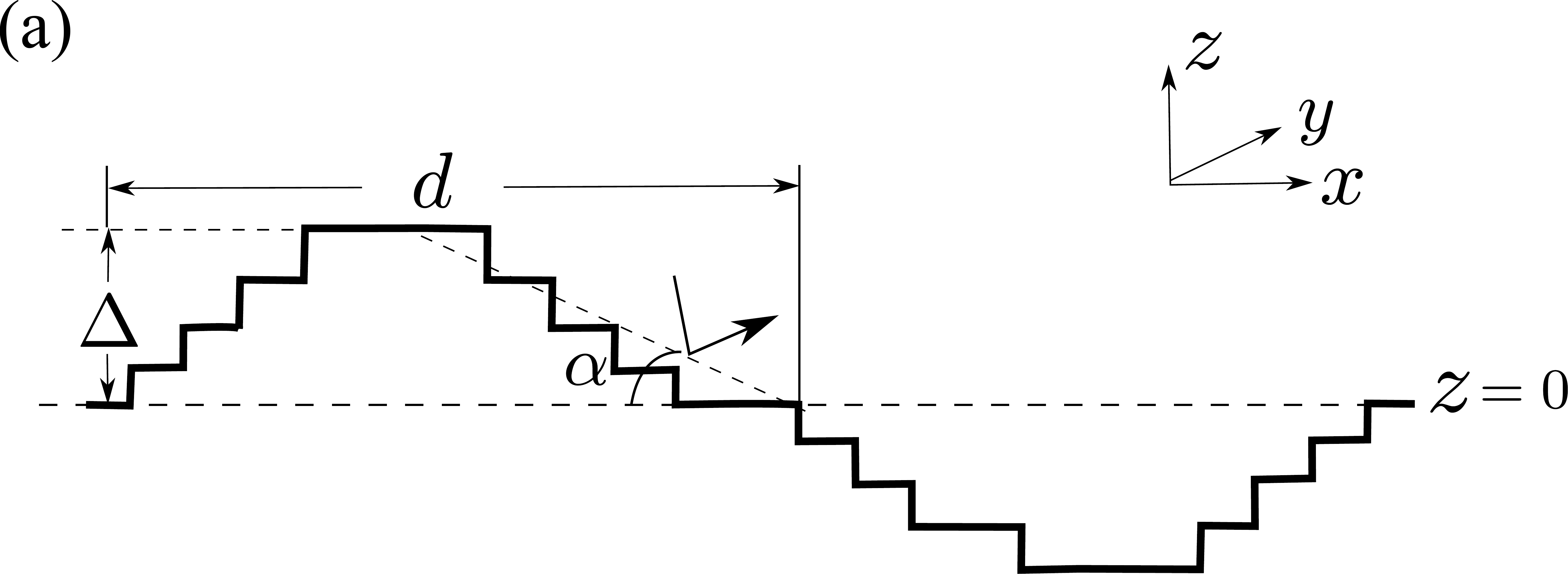}\\
\\
\\
\includegraphics[width=0.47\textwidth]{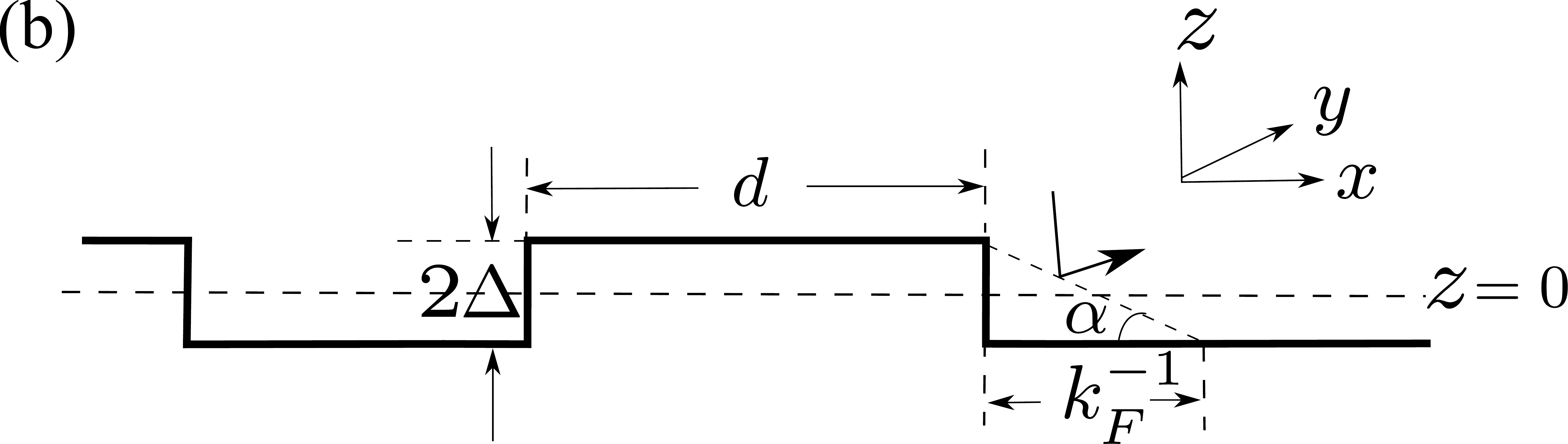}
\end{array}$
\caption{Two types of surface roughness. (a) The Gaussian type of roughness. Here the lattice discreteness can be ignored where $\Delta\gg a$ and $a$ is the lattice constant. (b) The exponential kind of roughness. The size of the roughness is $d$ and the height fluctuates as $\pm\Delta$ with respect to the average interface ($z=0$) where $2\Delta=a$. When the electron wavelength $k_F^{-1}\gg d$, the felt height of roughness is averaged result $\Delta/\sqrt{1/k_F^2d^2}=\Delta k_Fd$ on a length scale of $k_F^{-1}$ for both types of surface roughness. At $k_F^{-1}\ll d$, the incident electron feels only a single hill/valley or island. For Gaussian roughness, the electron scatters on the slope determined by the angle $\alpha\sim \Delta/d$ shown in (a). For the exponential roughness, however, the electron is scattered by the island edge which has a height $\Delta$ and an effective length $k_F^{-1}$, and thus the effective angle is $\alpha\simeq\Delta k_F$ as shown in (b).}\label{fig:roughness}
\end{figure}
The surface roughness is a random shift of the interface $\Delta(\vec{r})$ from $z=0$ so that $<\Delta(\vec{r})>=0$, where $\vec{r}=(x,y)$ is the coordinate in $z=0$ interface plane (see Fig. \ref{fig:roughness}). The roughness is described by the height correlator and its Fourier transform
\begin{equation}\label{eq:corre}
\begin{aligned}
<\Delta(\vec{r})\Delta(\vec{r'})>=&W(\vec{r}-\vec{r'}),\\
<|\Delta(q)|^2>=&W(q).
\end{aligned}
\end{equation}
Two main models of roughness are used in literature. One is Gaussian
\begin{equation}\label{eq:Gauss}
\begin{aligned}
W(\vec{r}-\vec{r'})=&\Delta^2e^{-(\vec{r}-\vec{r'})^2/d^2},\\
W(q)=&\pi\Delta^2d^2e^{-q^2d^2/4},
\end{aligned}
\end{equation}
where $d,\,\Delta$ are the characteristic size and height of the roughness, $\Delta\ll d, \,L$. This is the model widely used in earlier studies such as for the single subband mobility \cite{Entin,Ando_1977,Ando_1979,Ando_1982,Suris,Sakaki,Gold}.

However, later experimental observations found that the spacial correlations are more likely to follow an exponential behavior \cite{Goodnick,Feenstra_1994}
\begin{equation}\label{eq:exp}
\begin{aligned}
W(\vec{r}-\vec{r'})=&\Delta^2e^{-\sqrt{2}|\vec{r}-\vec{r'}|/d},\\
W(q)=&\pi\Delta^2d^2(1+q^2d^2/2)^{-3/2}.
\end{aligned}
\end{equation}
The important difference from the Gaussian case is that here $W(q)$ decays much slower as $q^{-3}$ at large $q$. This leads to a stronger scattering at large $n$. One way to envision this kind of roughness is to think about randomly distributed flat islands of an additional lattice layer with typical size $d$ on the top of the last complete layer of the crystal. These islands may, for example occupy half of the surface area, so that $\Delta(\vec{r})= \pm \Delta$ appear with equal probability where $2\Delta = a$ is the lattice constant. Whenever two survey points $\vec{r}$ and $\vec{r'}$ fall within the same island, $\Delta(\vec{r})\Delta(\vec{r'})$ is $\Delta^{2}$. This typically happens when points are close, i.e., $|\vec{r}-\vec{r'}|\ll d$. When one of the points misses this island $\Delta(\vec{r})\Delta(\vec{r'})$ is $-\Delta^{2}$. The probability of falling into different islands at $|\vec{r}-\vec{r'}|\ll d$ is $\propto |\vec{r}-\vec{r'}|/d$. So $W(\vec{r}-\vec{r}')\sim \Delta^2(1-|\vec{r}-\vec{r'}|/d)$. Such a behavior at small distances determines the large $q$ asymptote of the Fourier transform $W(q)$ as $\sim 1/q^3$, which is the result obtained at large $q$ from Eq. \eqref{eq:exp}. Our calculations of the roughness-limited mobility for accumulation layers are focused on this type of surface roughness. However, to make a connection with earlier studies\cite{Entin,Ando_1977,Ando_1979,Ando_1982,Suris,Sakaki,Gold}, we will also calculate the mobility for the Gaussian model and compare the results of these two models.

\section{Quasi-classical picture}\label{sec:class}
Inspired by Ref. \onlinecite{Suris}, in this section we start from an intuitive quasi-classical picture of the electron scattering by the surface roughness and get the scaling result shown in Fig. \ref{fig:conductivity_the_result}.

Electrons are scattered when they hit the rough ``hard wall" surface. The time between two consecutive collisions of electrons with the surface is $\sim L/v_F\sim m^*L/\hbar k_F$. For each bounce, the reflection is specular with respect to the tangential plane of the hitting point and therefore adds a random angle $\alpha$ to the direction of the reflected momentum. Due to this angular diffusion, the total relaxation of the momentum direction requires $\alpha^{-2}$ times collisions. Thus the relaxation time is
\begin{equation}\label{eq:relaxation time}
\tau=\frac{m^*L}{\hbar k_F\alpha^2}.
\end{equation}
Below we are going to investigate the deviation angle $\alpha$ at different values of $k_F$ and thus find $\tau$.

For the exponential surface roughness, one can imagine the irregularities as islands going up or down. Each island is flat on a scale $d$ and drops or rises abruptly by a height $\Delta$ on the edges. An electron can be regarded as a particle only on length scales larger than the wavelength $ k_F^{-1}$. At $d\ll k_F^{-1}$ or $1/d\gg k_F$, electrons can only feel an averaged roughness of all islands within the region of size $k_F^{-1}$ whose number is $(k_F^{-1})^2/d^2=1/k_F^2d^2$. Due to the randomness of the distribution of these islands, the resulting height or depth has a magnitude $\sim \Delta/\sqrt{1/k_F^2d^2}=\Delta k_Fd$. Such a height/depth on a length scale $k_F^{-1}$ effectively results in $\alpha=\Delta k_Fd/ k_F^{-1}=\Delta dk_F^2$. Using Eq. \eqref{eq:relaxation time}, we get for $k_Fd\ll 1$
\begin{equation}\label{eq:tau_small_d}
\tau\sim\frac{1/(\Delta d k_F^2)^2}{\hbar k_F/m^*L}=\frac{m^*L}{\hbar\Delta^2d^2k_F^5}.
\end{equation}
In the opposite case at $k_F^{-1}\ll d$, the electron hits a single island each time it bounces off the surface. However, when electrons hit the flat middle plane of the island, there is no momentum relaxation. Only when electrons happen to hit the sharp edges can they get a ``scattering" reflection. Since only on a scale $k_F^{-1}$ can electrons be seen as quasi-classical particles, the scattering edge is then estimated to be of height $\Delta$ and size $k_F^{-1}$ which gives rise to $\alpha=\Delta k_F$ (see Fig. \ref{fig:roughness}b). The probability to hit one such edge is proportional to its area fraction $k_F^{-1}d/d^2=1/k_Fd$. This gives
\begin{equation}\label{eq:tau_large d}
\tau\sim\frac{1/(\Delta k_F)^2}{1/k_Fd}\frac{m^*L}{\hbar k_F}=\frac{m^*Ld}{\hbar\Delta^2k_F^2}.
\end{equation}
Since $L\sim a_B/(na_B^2)^{1/5},$ $k_F\sim (na_B^2)^{2/5}/a_B$ as we said in Sec. \ref{sec:accu}, we get the mobility as
\begin{equation}\label{eq:mobility}
\mu\sim \frac{e}{h}\frac{1}{\Delta^2}\times\left\{ \begin{array}{ll}
\displaystyle{\frac{a_B^{8/5}}{d^2 n^{11/5} }}, &na_B^2\ll(a_B/d)^{5/2},\\
&\\
\displaystyle{\frac{a_Bd}{n}} \,\,\,,& na_B^2\gg (a_B/d)^{5/2},
\end{array} \right.
\end{equation}
and the 2D conductivity $\sigma=ne\mu$ as shown in Fig. \ref{fig:conductivity_the_result}, where $k_Fd\sim 1$ at $na_B^2\sim (a_B/d)^{5/2}$. When $d>a_B$, in the 3D regime where $na_B^2>1$ there is no range of $k_Fd\ll 1$ and the mobility always decreases as $\propto 1/n$. One should note that the 2D conductivity saturation $\sigma/(2e^2/h)\sim da_B/\Delta^2$ at large concentration $n$ is usually much larger than unity as $d,\,a_B\gg \Delta$ and implies that the accumulation layer remains metallic at large concentrations.

\section{Single subband case}\label{sec:sing}
In previous section, we have employed a quasi-classical perspective to understand the electron scattering off the surface roughness. Now we turn to the more formal quantum-mechanical approach.

Let us start from the single subband case where the scattering occurs within the same subband. The scattering rate $1/\tau$ of an electron at the Fermi level with the wave vector $\vec{k'}$ can be found according to Fermi's golden rule:
\begin{equation}
\frac{1}{\tau}=\frac{2\pi}{\hbar}\int \frac{d^2\vec{k}}{(2\pi)^2} \frac{|U(q)|^2}{\epsilon(q)^2}\delta(\varepsilon-\varepsilon_F)(1-\cos\theta')
\end{equation}
where $\varepsilon=\hbar^2k^2/2m^*$, $\varepsilon_F=\hbar^2k'^2/2m^*=\hbar^2k_F^2/2m^*$ are the final and initial (Fermi level) energies of an electron, $\vec{k}$ is the final electron momentum, $k_F$ is the Fermi wavenumber, $\theta'$ is the angle between initial and final electron momenta and $q=2k_F\sin(\theta'/2)$ is the magnitude of the transferred momentum $\vec{q}=\vec{k}-\vec{k'}$. All the momenta here are in the $x$-$y$ plane. Due to the electronic screening inside a single subband, the Fourier transform of the scattering potential $U(q)$ is reduced by the dielectric function $\epsilon(q)$ \cite{Ando_1982}
\begin{equation}\label{eq:2D_screening}
\epsilon(q)\simeq 1+2/a_Bq.
\end{equation}
Let us first derive the scattering potential resulting from the surface roughness. We know that electrons are confined near the interface and thus have a quantization kinetic energy $E_z=\hbar^2k_z^2/2m^*$ where $k_z$ is a multiple of $\pi/L$ (for the first subband, $k_z=\pi/L$). Due to the surface roughness, the confinement width $L$ fluctuates by $\Delta(\vec{r})$ at position $r$. The kinetic energy then varies by $(dE_z/dL)\Delta(\vec{r})\sim E_z\Delta(\vec{r})/L$. These fluctuations of the quantization kinetic energy act as a fluctuating potential $U(\vec{r})$ for the 2-dimensional motion of confined electrons. Its scattering matrix element for 2D Bloch states $U(q)$ within a given subband then satisfies
\begin{equation}\label{eq:scatter_matrix_element}
|U(q)|^2= \left(\frac{E_z}{L}\right)^2W(q) .
\end{equation}
As a result we get:
\begin{equation}\label{eq:quantum_large_d}
\frac{1}{\tau}\sim\frac{\hbar}{m^*}\frac{k_z^4}{L^2}\int d\theta' \frac{W(q)}{\epsilon(q)^2}(1-\cos\theta').
\end{equation}

At $k_Fd\ll 1$, according to Eqs. \eqref{eq:Gauss} and \eqref{eq:exp},  two models of roughness give the same $W(q)\sim \Delta^2d^2$. For the one subband case, $k_Fa_B \simeq 1$ and $\epsilon(q) \simeq 1$.
This gives the scattering rate as
\begin{equation}\label{eq:quantum_small_d}
\frac{1}{\tau}\sim\frac{\hbar}{m^*}\frac{k_z^4\Delta^2d^2}{L^2}.
\end{equation}
As $k_z\sim 1/L$, the mobility is then
\begin{equation}\label{eq:2D_mobility_small_d}
\mu\sim\frac{e}{\hbar}\frac{L^6}{\Delta^2d^2}.
\end{equation}
Since in the single subband case $na_B^2 \leq 1$, the condition of validity of Eq. \eqref{eq:2D_mobility_small_d} $k_Fd\ll 1$ is fulfilled for the roughness with $d < a_B$. The case $d < a_B$ was studied for silicon inversion layers in Ref. \onlinecite{Ando_1977} and discussed in Introduction above. For the 2D inversion layers, the width $L$ is determined by the applied electric field as $L\propto E^{-1/3}$ and one gets \cite{Ando_1977, Entin,Suris} $\mu\propto 1/E^2$. As the electron concentration increases, the interfacial electric field $E\propto n$ and mobility $\mu\propto 1/n^{2}$. Such a dependence was obtained in Ref. \onlinecite{Ando_1977} and used in Ref.~\onlinecite{DasSarma} for extrapolation to the multisubband case as shown in Fig. \ref{fig:conductivity_the_result}.
Note that for the 2D quantum wells, the width $L$ of the electron gas is the same as the well width so that Eq. \eqref{eq:2D_mobility_small_d} agrees with the well known result of Ref. \onlinecite{Sakaki}.

However, if $d>a_B$, for the single subband case, there is also a range of concentrations that satisfies $k_Fd\gg 1$. The results of two roughness models are different. 
For the Gaussian case, the typical $q$ is $\sim 1/d$ and the typical $\theta'$ is $\sim q/k_F\sim 1/k_Fd$, $W(q)\sim \Delta^2d^2$, $\epsilon(q)\sim 1+d/a_B\sim d/a_B$.  The scattering rate is then
\begin{equation}\label{eq:quantum_large_d_Gauss_screen}
\frac{1}{\tau}\sim \frac{\hbar}{m^*}\frac{k_z^4\Delta^2a_B^2}{L^2k_F^3d^3}.
\end{equation}
Putting $k_z\sim 1/L$ for the single subband case, we get the mobility as
\begin{equation}\label{eq:2D_mobility_large_d_Gauss}
\mu\sim\frac{e}{\hbar}\frac{L^6d^3k_F^3}{\Delta^2a_B^2},
\end{equation}
which can also be obtained from results in Refs. \onlinecite{Gold,Suris}.

For the exponential case, $W(q)$ decays in a much milder way as $\propto 1/q^3$ at large $q$. This leads to the large angle scattering. Indeed, let us consider the contribution to the integral in Eq. \eqref{eq:quantum_large_d} from the small angles $\theta' \sim (k_Fd)^{-1}$ and large angles $\theta' \sim 1$. In the first case, $W(q\sim 1/d)\sim\Delta^2d^2$, $\epsilon(q)\sim d/a_B$ and
$$\int_0^{(k_Fd)^{-1}}d\theta'(1-\cos\theta') \simeq \theta'^3 \simeq \frac{1}{(k_Fd)^3}.$$
As a result the contribution from the small angles to the scattering rate is the same as Eq. \eqref{eq:quantum_large_d_Gauss_screen}.
For the large angles  $W(q\sim k_F) \simeq \Delta^2 d^2 \left(k_Fd \right)^{-3}$ is $(k_Fd)^3$ times smaller than that of the small-angle scattering, $\epsilon(q)\simeq 1$ and the integral over the angle is
$$\int \limits_{(k_Fd)^{-1}}^{\pi} d\theta'(1-\cos\theta')  \sim 1.$$ The angle integral is $(k_Fd)^3$ times larger than that from the small angles, and the scattering rate is then
\begin{equation}\label{eq:quantum_large_d_exp}
\frac{1}{\tau}\sim\frac{\hbar}{m^*}\frac{k_z^4\Delta^2  }{L^2k_F^3 d},
\end{equation}
which due to the absence of screening is $(d/a_B)^2$ times larger than that from the small-angle scattering.
In this sense, the large-angle scattering is more effective and always gives the mobility as
\begin{equation}\label{eq:2D_mobility_large_d_exp}
\mu \sim \frac{e}{\hbar} \frac{L^6dk_F^3}{\Delta^2 }.
\end{equation}
The dominance of the large-angle scattering is a unique feature of the exponential roughness.

\section{Intersubband scattering in multisubband accumulation layers}\label{sec:inter}
In Sec. \ref{sec:sing} we have calculated the mobility limited by the surface roughness scattering of a single subband. We not only have recovered the results for the Gaussian type of roughness obtained by previous studies but also have got the results for the relatively unexplored exponential roughness. For multisubband accumulation layers, the situation is different from the single subband case. First, $k_z$ is not $\sim1/L$ but typically is $\sim k_F$, where $k_F=(n/L)^{1/3}$ is the 3D Fermi wavenumber of the electron gas. Second, though the screening now is still two-dimensional \footnote{One could worry about the use of 2D screening for the 3D accumulation layer with width $L$. Indeed, it is known that in a metallic film with width $L$ at the distance $q^{-1}\gg L$ the screening is two-dimensional, while at $ q^{-1} \ll L$ it becomes three-dimensional. In our case, the 3D screening radius inside the layer is $r_D=a_B/(Na_B^3)^{1/6}\sim a_B/(na_B^2)^{1/5}\sim L .$ Thus, for $q^{-1}< L$ the 3D dielectric function is $\sim 1$. Therefore, the 3D screening may be ignored.}, the screening radius is $L$ instead of $a_B$ so that $\epsilon(q)\simeq 1+1/Lq$ due to collective screening of multiple subbands \footnote{The 2D screening radius $r_2$ here is not $a_B$ but $L$ because when deriving $r_2\propto d(e\varphi)/dn$ where $e\varphi$ characterizes the electron chemical potential for each subband, the total 2D concentration $n$ is the electron concentration within each subband times the number of subbands $k_FL$. Therefore, the 2D screening radius is reduced from $a_B$ to $a_B/k_FL\simeq a_B/(na_B^2)^{1/5}\sim L$.}. Last, in addition to the intrasubband scattering, there is also intersubband scattering.

Typically, the intersubband scattering rate is of the same order of the intrasubband scattering as shown in Appendix \ref{App:AppendixA}. Therefore, the final scattering rate is approximately the product of the typical intrasubband scattering rate and the total number of subbands which the initial electron can be scattered into. It is easy to check that the typical transferred momentum in the $z$-direction is of the same order as the typical transferred momentum $q$ in the $x$-$y$ plane. At $1/d\gg k_F$ where $q\sim k_F$, all subbands ``communicate" with each other and the total number is $k_FL$. Multiplying by $k_FL$ the intrasubband result Eq. \eqref{eq:quantum_small_d} with $k_z\sim k_F$, we arrive at the final scattering rate given by Eq. \eqref{eq:tau_small_d}.
This is a universal result for both Gaussian and exponential models.
At $1/d\ll k_F$, for the exponential roughness, the typical transferred momentum is $q\sim 1/d$ for the small-angle scattering and $q\sim k_F$ for the large-angle one. So for the former, the number of subbands involved in the scattering process is $\sim L/d$, much smaller than the number $ k_FL$ for the latter. Considering also its immunity to screening as $k_FL\gg1$, the large-angle scattering mechanism absolutely dominates. Using the intrasubband scattering rate as given by Eq. \eqref{eq:quantum_large_d_exp} and $k_z\sim k_F$, we arrive at the total scattering rate at $k_Fd\gg 1$ for the exponential roughness given by Eq. \eqref{eq:tau_large d}. Thereby, we get the same expressions for the mobility as in Eq. \eqref{eq:mobility}. The corresponding 2D conductivity at large $n$ saturates as $\sigma/(2e^2/h)\sim da_B/\Delta^2$ (see Fig. \ref{fig:conductivity_the_result}), which means that there is no re-entrant metal-insulator transition. In Appendix \ref{App:AppendixA}, we do more careful estimation of the numerical coefficient in front of $da_B/\Delta^2$ and get that it is close to 1.

Compared to the simple result in the exponential case, the mobility for the Gaussian roughness is more complicated. At $k_F^{-1}\ll d<L$, the intrasubband scattering is unscreened for the Gaussian case and the rate is the same as Eq. \eqref{eq:quantum_large_d_exp}. The typical momentum transfer is $q\sim 1/d$, so only a few subbands the number of which is $\sim q/L^{-1}=L/d$ can participate in the intersubband scattering. The total scattering rate is then the result in Eq. \eqref{eq:quantum_large_d_exp} times $L/d$. At even larger $ d\gg L\gg k_F^{-1}$, the typical momentum transferred $1/d$ is smaller than the $z$-direction momentum quantization $1/L$. Therefore no intersubband scattering is possible. This situation resembles the single subband case and is natural since when $L\ll d$ we actually are dealing with a 2D system. At $d\gg L$, the screening of the potential adds the factor $(L/a_B)^2$ to the total scattering rate given by Eq. \eqref{eq:quantum_large_d_Gauss_screen} as the screening radius now is $L$ instead of $a_B$. The result is summarized as follows
\begin{numcases}
{\frac{1}{\tau}\sim }
\frac{\hbar k_F \Delta^2}{m^*Ld^2 }, & $k_F^{-1}\ll d\ll L\,,$\label{eq:3D_scattering_rate_large_d_Gauss_1}\\ \nonumber
&\\
\frac{\hbar k_F\Delta^2}{m^*d^3},& $L\ll d\,.$\label{eq:3D_scattering_rate_large_d_Gauss_2}\\ \nonumber
\end{numcases}
$k_F^{-1}=d$ is reached at $na_B^2\sim (a_B/d)^{5/2}$ and $d=L$ is achieved at $na_B^2\sim (a_B/d)^5$. By expressing $k_F$ and $L$ in terms of $n$, one can then get the mobility as a function of the 2D electron concentration. The corresponding results together with that for the exponential case are listed in Table. \ref{tab:1}.
\begin{table}
\caption{\label{tab:1} Mobility $\mu$ in units of $\left(e/\hbar\right)\left(d^4/\Delta^2\right)$ as a function of the 2D electron concentration $n$ at different values of $d$ for two types of surface roughness, the Gaussian model (G) and the exponential one (E). Since the Fermi wavenumber $k_F\simeq (n/L)^{1/3}$, and the width of the 2D electron gas $L\sim a_B/(na_B^2)^{1/5}$, we get $k_F^{-1}=d$ at $na_B^2\sim (a_B/d)^{5/2}$ and $L=d$ at $na_B^2\sim(a_B/d)^5$. }
\begin{ruledtabular}
\renewcommand{\arraystretch}{2}
\begin{tabular}{ c  | c| c| c}
 & $d<k_F^{-1}$ & $k_F^{-1}<d<L$& $L<d$ \\ \hline
G$\quad$&$ (a_B/d)^6/(na_B^2)^{11/5}$&$(a_B/d)^{2}/(na_B^2)^{3/5}$&$(a_B/d)/(na_B^2)^{2/5}$\\ \hline
E$\quad$&$(a_B/d)^6/(na_B^2)^{11/5}$&$(a_B/d)^3/(na_B^2)$&$(a_B/d)^3/(na_B^2)$\\
\end{tabular}
\end{ruledtabular}
\end{table}

\begin{figure}[h]
\includegraphics[width=.45\textwidth]{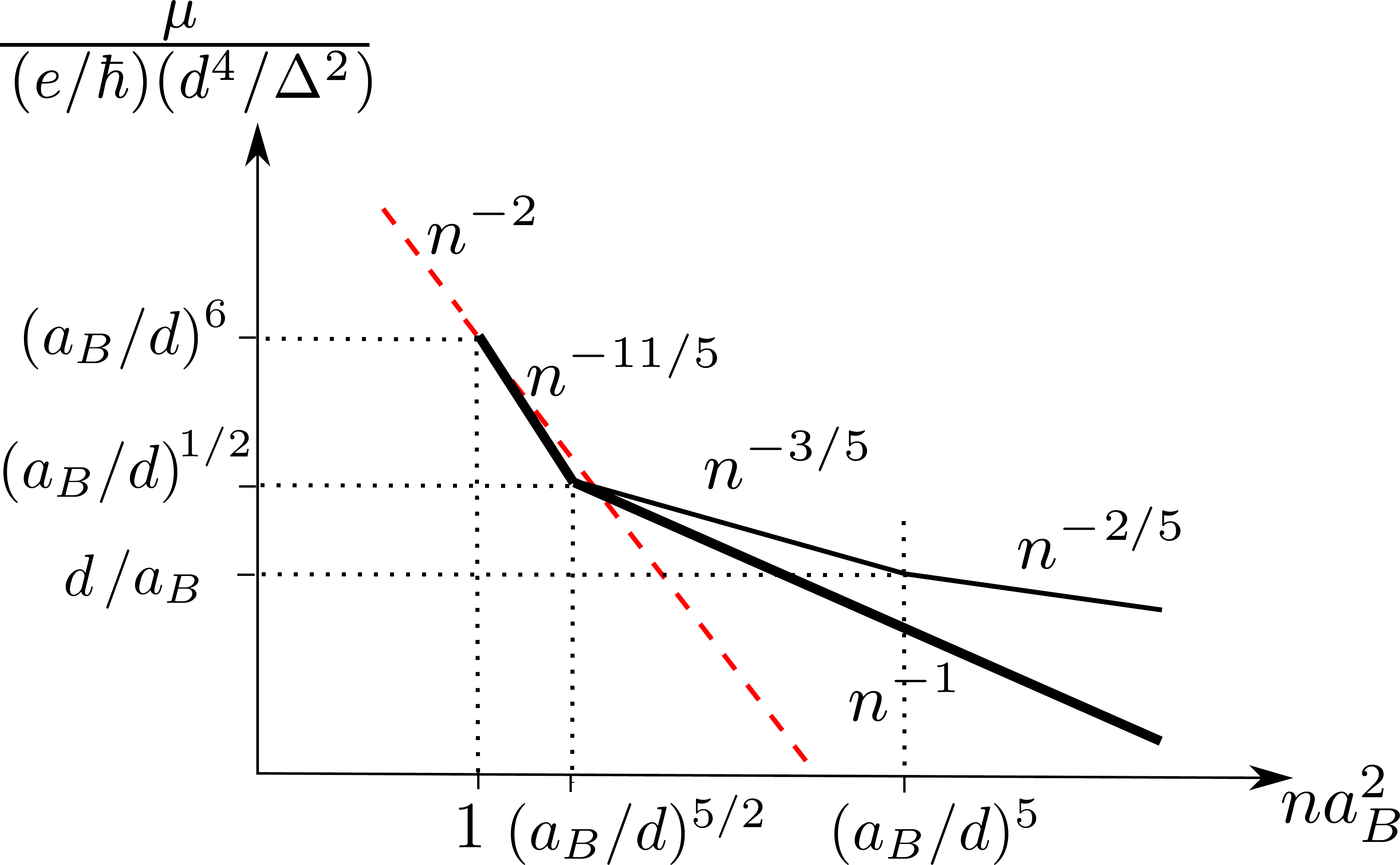}\\
\caption{The scaling behavior of the mobility $\mu$ in units of $\left(e/\hbar\right)\left(d^4/\Delta^2\right)$ as a function of the scaled electron concentration $na_B^{2}$ plotted at $d<a_B$ in a double logarithmic scale. The thick solid line (black) denotes the mobility of the accumulation layer for the exponential roughness. The thin solid line (black) represents the mobility for the Gaussian roughness which also decreases. Here only powers of the $n$ dependence are shown while the complete scaling formulae are presented in Table. \ref{tab:1}. The thin dashed line (red) represents the $1/n^2$ dependence derived for a single subband \cite{Ando_1977} and its conjectured extrapolation \cite{DasSarma} to larger concentrations. }\label{fig:mobility}
\end{figure}

The obtained $\mu(n)$ dependence is presented in Fig. \ref{fig:mobility}. For the exponential roughness, the corresponding 2D conductivity $\sigma$ is shown in Fig. \ref{fig:conductivity_the_result}. The smallest conductivity for the exponential case is larger than $2e^2/h$ as mentioned above. For the Gaussian model, the smallest $\sigma/(2e^2/h)$ is also $\sim da_B/\Delta^2$. Since in reality, $d>\Delta$, we get $\sigma/(2e^2/h)\gg1$. So, the smallest conductivity for the Gaussian roughness is also always above the critical value and no re-entrant metal-insulator transition will happen in realistic situations.

In Sec. \ref{sec:class}, we gave a quasi-classical explanation of the mobility limited by the exponential surface roughness. Inspired by Ref. \onlinecite{Suris}, we can interpret the Gaussian roughness results quasi-classically as well. Below we again start from Eq. \eqref{eq:relaxation time} and find $\alpha$ for the Gaussian roughness in different situations.
At $k_Fd\ll1$, the roughness relief is averaged over the electron wavelength and the resulting relaxation time is the same as in the exponential case given by Eq. \eqref{eq:tau_small_d}.
At $k_F^{-1}\ll d\ll L$, the wavelength $k_F^{-1}$ is smaller than the size of the roughness hill. The electron then collides with a single hill (valley) each time it hits the surface and the deviation angle $\alpha$ is the slope of each single hill (valley) $\sim \Delta/d$ (see Fig. \ref{fig:roughness}a). The relaxation time is given by Eq. \eqref{eq:3D_scattering_rate_large_d_Gauss_1}.
At $L\ll d$, the electronic screening changes the scattering potential by a factor $L/d$, which can effectively be regarded as reducing the roughness height from $\Delta$ to $\Delta L/d$. The size of each single hill is so large that the electron can hit the same hill consecutively for several times during which it has traveled back and forth for $\sim d/L$ times within the accumulation layer. Since these consecutive hits are on the same slope, the scattered angle is the same and $\alpha$ accumulates, in contrast with uncorrelated random collisions on different hills (valleys). After the electron finishes colliding with the same slope, the accumulated angle is $(\Delta L/d^2)(d/L)=(\Delta/d)\ll1$ and the time of such a series of collisions is $\sim(L/k_F)( d/L)=d/k_F$. The resulting relaxation time is then given by Eq. \eqref{eq:3D_scattering_rate_large_d_Gauss_2}.
Thus we obtained quasi-classically the same scaling behavior of the mobility limited by the Gaussian roughness as by the quantum-mechanical approach.

\section{Conclusion}\label{sec:con}
In this paper, we have studied the surface-roughness limited mobility in inversion and multisubband accumulation layers as a function of the 2D electron concentration $n$
for two models of the surface roughness both quantum-mechanically and quasi-classically. For the more realistic exponential roughness, the mobility decreases as $\propto 1/n$ at large $n$ and results in a 2D conductivity saturation as $\sigma/(2e^2/h)\simeq d a_B/\Delta^2\gg 1$ since the characteristic roughness size $d$ and the effective Bohr radius $a_B$ are larger than the characteristic roughness height $\Delta\simeq a/2$ where $a$ is the lattice constant. For the Gaussian roughness which was widely used in earlier studies, the minimum conductivity is found to be larger than the critical value as well. One should note that the considerations here have not included the contribution from the tail electrons, which makes $\sigma$ even larger in reality due to their larger distances from the surface and thus larger relaxation times \cite{RAT}.
So there is no reason to expect the re-entrant metal-insulator transition~\cite{DasSarma} at large concentrations. Indeed, decent conductivities were observed in large concentration accumulation layers in Refs. \onlinecite{diamond,Iwasa_2009,Iwasa_2012,Tardella}. 

$\phantom{}$
\vspace*{2ex} \par \noindent
{\em Acknowledgments.}

We are grateful to S. A. Campbell, A. V. Chaplik, M. V. Entin, B. Jalan, M. J. Manfra, M. Sammon, M. Shur, and R. A. Suris for helpful discussions.
This work was supported primarily by the National Science Foundation through the University of Minnesota MRSEC under Award No. DMR-1420013.

\appendix

\section{Numerical coefficients in $\mathbf{\sigma(n)}$ dependence in the exponential model} \label{App:AppendixA}

Eq. \eqref{eq:concentration_linear} shows that about $90\%$ of electrons are located within a distance $L/2$ from the interface. So it is a good approximation to assume that electrons inside the accumulation layer are confined within a width $D=L/2$. Then, the electron wave function of each subband is
\begin{equation}\label{eq:wave_function}
\xi(r,z)\simeq \sqrt{\frac{2}{D}}\exp(i\vec{k_r}\cdot\vec{ r})\sin(k_z z)
\end{equation}
where $\vec{k_r}$, $k_z =m\pi /D$ with $m$ being a positive integer are respectively the $x$-$y$ plane and $z$-direction momenta of electrons in this subband and different subbands correspond to different values of $k_z$.
Therefore, similar to that in Ref. \onlinecite{Ando_1977}, the matrix element $U(q)$ satisfies
\begin{equation}\label{potential}
<|U|^2>=\frac{8}{D}\varepsilon_F^2\left(\frac{k_{z}^{'}k_z}{k_F^2}\right)^2W(q)
\end{equation}
where an isotropic mass spectrum is assumed, $k_{z}^{'},\,k_{z}$ are the initial and final $z$-components of the electron wavevector, $q=|\vec{q}|$ and $\vec{q}$ is the momentum transferred in the $x$-$y$ plane.

Since $|U|^2$ is isotropic only with respect to the 2D momentum $\vec{q}$ in the $x$-$y$ plane instead of the total 3D transferred momentum, we use the more general expression for the scattering rate as can be obtained from Ref. \onlinecite{Siggia}
\begin{equation}\label{eq:scattering_rate}
\frac{1}{\tau}=\frac{2\pi}{\hbar}\int \frac{d^3\vec{k}}{(2\pi)^3} \frac{|U|^2}{\epsilon(q)^2}\delta(\varepsilon-\varepsilon_F)\left(1-\frac{k_r\cos\phi}{k_r^{'}}\right),
\end{equation}
where $k_r=|\vec{k_r}|,\,k_r'=|\vec{k_r'}|$, and $\vec{k_r},\,\vec{k_r'}$ are the $x$-$y$ components of the final and initial momenta $\vec{k}$ and $\vec{k'}$, $\phi $ is the angle between $\vec{k_r}$ and $\vec{k_r'}$. (Here we assume a constant relaxation time for different subbands, which is a good approximation for electrons located within the distance $D$.)

At $k_Fd\ll 1$, for both models of roughness, $W(q)=\pi \Delta^2d^2,\,\epsilon(q)\simeq 1$.
So
\begin{equation}
\begin{aligned}
\frac{1}{\tau}&=\frac{2}{\pi \hbar}\frac{\varepsilon_F^2\Delta^2d^2}{D}\int k_rdk_rd\phi dk_z\left(\frac{k_{z}^{'}k_z}{k_F^2}\right)^2\\
&\quad\left(1-\frac{k_r\cos\phi}{k_r^{'}}\right)\delta\left[\frac{\hbar^2k_F^2}{2m^*}-\frac{\hbar^2(k_z^2+k_r^2)}{2m^*}\right]\\
&=\frac{2\varepsilon_F\Delta^2d^2k_F^3}{\hbar D}\int d(\cos\theta)\left(\cos\theta\cos\theta_0\right)^2\\
\end{aligned}
\end{equation}
where $\cos\theta=k_z/k_F,\,\cos\theta_0=k_z^{'}/k_F$.
One should note that this scattering rate is for one specific direction of $k'$. To get the averaged result, one should average over all $\theta_0$ and arrive at
\begin{equation}\label{small_d}
\begin{aligned}
\frac{1}{\tau}\simeq\frac{0.2\varepsilon_F\Delta^2d^2k_F^3}{\hbar D}.
\end{aligned}
\end{equation}

At $k_F d\gg 1$, for the exponential case, again we have $\epsilon(q)\simeq 1$. By using Eq. \eqref{eq:exp}, we get
\begin{equation}\label{exp_large_d}
\begin{aligned}
\frac{1}{\tau}=&\frac{2}{\pi \hbar}\frac{\varepsilon_F^2\Delta^2d^2}{D}\int k_rdk_rd\phi dk_zd(k_z^{'}/k_F)\left(\frac{k_{z}^{'}k_z}{k_F^2}\right)^2\\
&\quad\,\, \left\{1+\left[k_r^2+(k_{r}^{'})^2-2k_rk_{r}^{'}\cos\phi\right]d^2/2\right\}^{-3/2}\\
&\quad\left(1-\frac{k_{r}\cos\phi}{k_r^{'}}\right)\delta\left[\frac{\hbar^2k_F^2}{2m^*}-\frac{\hbar^2(k_z^2+k_r^2)}{2m^*}\right]\\
\simeq&\frac{1.5\varepsilon_F\Delta^2}{\hbar Dd},
\end{aligned}
\end{equation}
which matches Eq. \eqref{small_d} at $k_Fd\simeq 2$. Substituting $k_F=(3\pi^2n/D)^{1/3}$ into Eqs. \eqref{small_d} and \eqref{exp_large_d}, where $D=L/2$ and $L$ is given by Eq. \eqref{eq:decay}, we obtain numerical coefficients for $\sigma(n)$ and get the corresponding saturation value $\sigma/(2e^2/h)\simeq 0.6 da_B/\Delta^2$ at the point $na_B^2\simeq 0.4 (a_B/d)^{5/2}$.

\bibliography{roughness}

\end{document}